\documentclass[aps,prl,twocolumn,reprint,superscriptaddress, floatfix]{revtex4-2}
\usepackage{graphicx,amsfonts,color,comment,amsmath,hyperref,float}
\usepackage{dcolumn}
\usepackage{bm}
\usepackage{url}
\usepackage{soul}
\begin{document}

\title{Extended TeV halos may commonly exist around middle-aged pulsars}
\author{A.~Albert}
\affiliation{Los Alamos National Laboratory, Los Alamos, NM, USA }

\author{R.~Alfaro}
\affiliation{Instituto de Física, Universidad Nacional Autónoma de México, Ciudad de Mexico, Mexico }

\author{C.~Alvarez}
\affiliation{Universidad Autónoma de Chiapas, Tuxtla Gutiérrez, Chiapas, México}

\author{J.C.~Arteaga-Velázquez}
\affiliation{Universidad Michoacana de San Nicolás de Hidalgo, Morelia, Mexico }

\author{D.~Avila Rojas}
\affiliation{Instituto de Física, Universidad Nacional Autónoma de México, Ciudad de Mexico, Mexico }

\author{H.A.~Ayala Solares}
\affiliation{Department of Physics, Pennsylvania State University, University Park, PA, USA }

\author{R.~Babu}
\affiliation{Department of Physics, Michigan Technological University, Houghton, MI, USA }

\author{E.~Belmont-Moreno}
\affiliation{Instituto de Física, Universidad Nacional Autónoma de México, Ciudad de Mexico, Mexico }

\author{A.~Bernal}
\affiliation{Instituto de Astronomía, Universidad Nacional Autónoma de México, Ciudad de Mexico, Mexico }

\author{K.S.~Caballero-Mora}
\affiliation{Universidad Autónoma de Chiapas, Tuxtla Gutiérrez, Chiapas, México}

\author{T.~Capistrán}
\affiliation{Università degli Studi di Torino, I-10125 Torino, Italy}

\author{A.~Carramiñana}
\affiliation{Instituto Nacional de Astrofísica, Óptica y Electrónica, Puebla, Mexico }

\author{S.~Casanova}
\affiliation{Instytut Fizyki Jadrowej im Henryka Niewodniczanskiego Polskiej Akademii Nauk, IFJ-PAN, Krakow, Poland }

\author{U.~Cotti}
\affiliation{Universidad Michoacana de San Nicolás de Hidalgo, Morelia, Mexico }

\author{J.~Cotzomi}
\affiliation{Facultad de Ciencias Físico Matemáticas, Benemérita Universidad Autónoma de Puebla, Puebla, Mexico }

\author{S.~Coutiño de León} 
\affiliation{Department of Physics, University of Wisconsin-Madison, Madison, WI, USA }
\affiliation{Instituto de Física Corpuscular, CSIC, Universitat de València, E-46980, Paterna, Valencia, Spain}
\affiliation{\href{mailto:scoutino@ific.uv.es}{scoutino@ific.uv.es}}

\author{E.~De la Fuente}
\affiliation{Departamento de Física, Centro Universitario de Ciencias Exactase Ingenierias, Universidad de Guadalajara, Guadalajara, Mexico }

\author{C.~de León}
\affiliation{Universidad Michoacana de San Nicolás de Hidalgo, Morelia, Mexico } 

\author{D.~Depaoli}
\affiliation{Max-Planck Institute for Nuclear Physics, 69117 Heidelberg, Germany}

\author{P.~Desiati}
\affiliation{Department of Physics, University of Wisconsin-Madison, Madison, WI, USA}

\author{N.~Di Lalla}
\affiliation{Department of Physics, Stanford University: Stanford, CA 94305–4060, USA}

\author{R.~Diaz Hernandez}
\affiliation{Instituto Nacional de Astrofísica, Óptica y Electrónica, Puebla, Mexico }

\author{B.L.~Dingus}
\affiliation{Los Alamos National Laboratory, Los Alamos, NM, USA }

\author{M.A.~DuVernois}
\affiliation{Department of Physics, University of Wisconsin-Madison, Madison, WI, USA }

\author{J.C.~Díaz-Vélez}
\affiliation{Department of Physics, University of Wisconsin-Madison, Madison, WI, USA }

\author{K.~Engel}
\affiliation{Department of Physics, University of Maryland, College Park, MD, USA }

\author{C.~Espinoza}
\affiliation{Instituto de Física, Universidad Nacional Autónoma de México, Ciudad de Mexico, Mexico }

\author{K.L.~Fan}
\affiliation{Department of Physics, University of Maryland, College Park, MD, USA }

\author{K.~Fang}
\affiliation{Department of Physics, University of Wisconsin-Madison, Madison, WI, USA }
\affiliation{\href{mailto:kefang@physics.wisc.edu}{kefang@physics.wisc.edu}}

\author{N.~Fraija}
\affiliation{Instituto de Astronomía, Universidad Nacional Autónoma de México, Ciudad de Mexico, Mexico }

\author{J.A.~García-González}
\affiliation{Tecnologico de Monterrey, Escuela de Ingeniería y Ciencias, Ave. Eugenio Garza Sada 2501, Monterrey, N.L., Mexico, 64849}

\author{F.~Garfias}
\affiliation{Instituto de Astronomía, Universidad Nacional Autónoma de México, Ciudad de Mexico, Mexico }

\author{H.~Goksu}
\affiliation{Max-Planck Institute for Nuclear Physics, 69117 Heidelberg, Germany}

\author{M.M.~González}
\affiliation{Instituto de Astronomía, Universidad Nacional Autónoma de México, Ciudad de Mexico, Mexico }

\author{J.A.~Goodman}
\affiliation{Department of Physics, University of Maryland, College Park, MD, USA }

\author{S.~Groetsch}
\affiliation{Department of Physics, Michigan Technological University, Houghton, MI, USA }

\author{J.P.~Harding}
\affiliation{Los Alamos National Laboratory, Los Alamos, NM, USA }

\author{S.~Hernández-Cadena}
\affiliation{Instituto de Física, Universidad Nacional Autónoma de México, Ciudad de Mexico, Mexico }

\author{I.~Herzog}
\affiliation{Department of Physics and Astronomy, Michigan State University, East Lansing, MI, USA }

\author{D.~Huang}
\affiliation{Department of Physics, University of Maryland, College Park, MD, USA }

\author{F.~Hueyotl-Zahuantitla}
\affiliation{Universidad Autónoma de Chiapas, Tuxtla Gutiérrez, Chiapas, México}

\author{A.~Iriarte}
\affiliation{Instituto de Astronomía, Universidad Nacional Autónoma de México, Ciudad de Mexico, Mexico }

\author{S.~Kaufmann}
\affiliation{Universidad Politecnica de Pachuca, Pachuca, Hgo, Mexico }

\author{D.~Kieda}
\affiliation{Department of Physics and Astronomy, University of Utah, Salt Lake City, UT, USA }

\author{J.~Lee}
\affiliation{University of Seoul, Seoul, Rep. of Korea}

\author{H.~León Vargas}
\affiliation{Instituto de Física, Universidad Nacional Autónoma de México, Ciudad de Mexico, Mexico }

\author{J.T.~Linnemann}
\affiliation{Department of Physics and Astronomy, Michigan State University, East Lansing, MI, USA }

\author{A.L.~Longinotti}
\affiliation{Instituto de Astronomía, Universidad Nacional Autónoma de México, Ciudad de Mexico, Mexico }

\author{G.~Luis-Raya}
\affiliation{Universidad Politecnica de Pachuca, Pachuca, Hgo, Mexico }

\author{K.~Malone}
\affiliation{Los Alamos National Laboratory, Los Alamos, NM, USA }

\author{O.~Martinez}
\affiliation{Facultad de Ciencias Físico Matemáticas, Benemérita Universidad Autónoma de Puebla, Puebla, Mexico }

\author{J.~Martínez-Castro}
\affiliation{Centro de Investigaci'on en Computaci'on, Instituto Polit'ecnico Nacional, M'exico City, M'exico.}

\author{J.A.~Matthews}
\affiliation{Dept of Physics and Astronomy, University of New Mexico, Albuquerque, NM, USA }

\author{P.~Miranda-Romagnoli}
\affiliation{Universidad Autónoma del Estado de Hidalgo, Pachuca, Mexico }

\author{J.A.~Morales-Soto}
\affiliation{Universidad Michoacana de San Nicolás de Hidalgo, Morelia, Mexico }

\author{E.~Moreno}
\affiliation{Facultad de Ciencias Físico Matemáticas, Benemérita Universidad Autónoma de Puebla, Puebla, Mexico }

\author{M.~Mostafá}
\affiliation{Department of Physics, Temple University, Philadelphia, PA, USA}

\author{L.~Nellen}
\affiliation{Instituto de Ciencias Nucleares, Universidad Nacional Autónoma de Mexico, Ciudad de Mexico, Mexico }

\author{M.U.~Nisa}
\affiliation{Department of Physics and Astronomy, Michigan State University, East Lansing, MI, USA }

\author{N.~Omodei}
\affiliation{Department of Physics, Stanford University: Stanford, CA 94305–4060, USA}

\author{Y.~Pérez Araujo}
\affiliation{Instituto de Física, Universidad Nacional Autónoma de México, Ciudad de Mexico, Mexico }

\author{E.G.~Pérez-Pérez}
\affiliation{Universidad Politecnica de Pachuca, Pachuca, Hgo, Mexico }

\author{C.D.~Rho}
\affiliation{Department of Physics, Sungkyunkwan University, Suwon 16419, South Korea}

\author{D.~Rosa-González}
\affiliation{Instituto Nacional de Astrofísica, Óptica y Electrónica, Puebla, Mexico }

\author{E.~Ruiz-Velasco}
\affiliation{Max-Planck Institute for Nuclear Physics, 69117 Heidelberg, Germany}

\author{H.~Salazar}
\affiliation{Facultad de Ciencias Físico Matemáticas, Benemérita Universidad Autónoma de Puebla, Puebla, Mexico }

\author{D.~Salazar-Gallegos}
\affiliation{Department of Physics and Astronomy, Michigan State University, East Lansing, MI, USA }

\author{A.~Sandoval}
\affiliation{Instituto de Física, Universidad Nacional Autónoma de México, Ciudad de Mexico, Mexico }

\author{M.~Schneider}
\affiliation{Department of Physics, University of Maryland, College Park, MD, USA }

\author{J.~Serna-Franco}
\affiliation{Instituto de Física, Universidad Nacional Autónoma de México, Ciudad de Mexico, Mexico }

\author{Y.~Son}
\affiliation{University of Seoul, Seoul, Rep. of Korea}

\author{R.W.~Springer}
\affiliation{Department of Physics and Astronomy, University of Utah, Salt Lake City, UT, USA }

\author{O.~Tibolla}
\affiliation{Universidad Politecnica de Pachuca, Pachuca, Hgo, Mexico }

\author{K.~Tollefson}
\affiliation{Department of Physics and Astronomy, Michigan State University, East Lansing, MI, USA }

\author{I.~Torres}
\affiliation{Instituto Nacional de Astrofísica, Óptica y Electrónica, Puebla, Mexico }

\author{R.~Torres-Escobedo}
\affiliation{Tsung-Dao Lee Institute, Shanghai Jiao Tong University, Shanghai, China}

\author{R.~Turner}
\affiliation{Department of Physics, Michigan Technological University, Houghton, MI, USA }

\author{F.~Ureña-Mena}
\affiliation{Instituto Nacional de Astrofísica, Óptica y Electrónica, Puebla, Mexico }

\author{E.~Varela}
\affiliation{Facultad de Ciencias Físico Matemáticas, Benemérita Universidad Autónoma de Puebla, Puebla, Mexico }

\author{L.~Villaseñor}
\affiliation{Facultad de Ciencias Físico Matemáticas, Benemérita Universidad Autónoma de Puebla, Puebla, Mexico }

\author{X.~Wang}
\affiliation{Department of Physics, Michigan Technological University, Houghton, MI, USA }

\author{E.~Willox}
\affiliation{Department of Physics, University of Maryland, College Park, MD, USA }

\author{H.~Wu}
\affiliation{Department of Physics, University of Wisconsin-Madison, Madison, WI, USA }
\affiliation{\href{mailto:hwu298@wisc.edu}{hwu298@wisc.edu}}
\author{H.~Zhou}
\affiliation{Tsung-Dao Lee Institute, Shanghai Jiao Tong University, Shanghai, China}
\date{\today}

\begin{abstract}
Extended gamma-ray emission around isolated pulsars at TeV energies, also known as TeV halos, have been found around a handful of middle-aged pulsars. The halos are significantly more extended than their pulsar wind nebulae but much smaller than the particle diffusion length in the interstellar medium.
The origin of TeV halos is unknown. Interpretations invoke either local effects related to the environment of a pulsar or generic particle transport behaviors. The latter scenario predicts that TeV halos would be a universal phenomena for all pulsars. We searched for extended gamma-ray emission around  36 isolated middle-aged pulsars identified by radio and gamma-ray facilities using 2321 days of data from the High-Altitude Water Cherenkov (HAWC) Observatory. Through a stacking analysis comparing TeV flux models against a background-only hypothesis, we identified TeV halo-like emission at a significance level of $5.10\,\sigma$. Our results imply that extended TeV gamma-ray halos may commonly exist around middle-aged pulsars. This reveals a previously unknown feature about pulsars and opens a new window to identify the pulsar population that is invisible to radio, x-ray, and GeV gamma-ray observations due to magnetospheric configurations.
\end{abstract}

\maketitle

Gamma-ray emission with angular size of a few degrees was first detected by the High-Altitude Water Cherenkov (HAWC) Observatory around two nearby middle-aged energetic pulsars between 8 and 40 TeV in 2017 \cite{gemingahawc17}. The TeV emission seen around those pulsars is much more extended than the pulsar wind nebulae whose sizes are no more than a few arc-minutes \cite{Posselt:2016lot, Birzan:2015puz, HESS_Geminga, 2019PhRvD.100l3015D}. These extended sources were named TeV halos \cite{Linden17}. Since the discovery of the first two TeV halos, more similar sources have been identified by HAWC \cite{3HWC} and LHAASO \cite{lhasso21}. To date, the online catalog for TeV Astronomy (TeVCat, {\url{http://tevcat.uchicago.edu}}) reports a total of about ten TeV halos and TeV halo candidates. Counterparts of TeV halos at lower energies have been found in {\it Fermi}-LAT \citep{DiMauro:2019yvh, 2024arXiv241107162A} and H.E.S.S. \citep{HESS:2023sbf} data.

The origin of the TeV halos is largely unknown \cite{Lopez-Coto:2022igd}. The emission cannot be explained by electron diffusion in the interstellar medium (ISM) as it requires a suppression of the cosmic-ray diffusivity by 100-1000 times \cite{gemingahawc17}. Several explanations have been proposed, but all have weaknesses. Models can be divided into two main classes, attributing the slow diffusion to a turbulence that universally exists around the central source \cite{2018PhRvD..98f3017E, 2022PhRvD.105l3008M} or pre-existing regions of non-standard  relativistic electron transport \cite{2019PhRvL.123v1103L, 2022PhRvD.106l3033D}. A related debate is whether the halos commonly exist around middle-aged pulsars \cite{2019PhRvD.100d3016S} or most middle-aged pulsars do not develop halos \cite{2022A&A...665A.132M}.

The several known TeV halos present similar physical extensions \cite{Linden17}, which are related to the cooling time of very-high-energy (0.1-100~TeV) electrons in the Cosmic Microwave Background (CMB) and interstellar radiation field. The flux of the TeV emission seems to be related to the pulsar's emissivity \cite{DiMauro2020}. Motivated by the similarity of the known sources, we perform the first systematic search for TeV halos in the HAWC data around pulsar populations that are identified by radio and GeV gamma-ray observations.

{\it Samples of Isolated Middle-aged Pulsars.---}Our analysis uses radio pulsars from the Australia Telescope National Facility (ATNF) catalog \footnote{\url{http://www.atnf.csiro.au/research/pulsar/psrcat} - version 1.71} \cite{2005AJ....129.1993M} and gamma-ray pulsars from the Third LAT Pulsar catalog (3PC) \cite{3PC}. We apply the following selection criteria: (1) a pulsar is in HAWC's field of view ($-26^{\circ} < \delta < 64^{\circ}$),  (2) not in a binary system, (3) middle-aged, defined as  the spin down age  $P/(2\,\dot{P})>20$ kyr, where $P$ and $\dot{P}$ are the barycentric period of the pulsar and its time derivative, respectively, and (4)  with the spin-down flux above the HAWC sensitivity. The spin-down flux is defined as $\dot{E}/(4\pi d^2)$, where $\dot{E}$ and $d$ are the spin-down power and distance, respectively.
A total of 1304 sources pass the first three selection criteria. 
Since the distance to a pulsar is needed to estimate its detectability and the angular size of its halo, we remove 3 pulsars without reported distances from the list. After applying the fourth criterion, the selection yields a total of 86 sources, including 49 gamma-ray pulsars and 72 radio-loud pulsars. Among the gamma-ray pulsars, 14 are radio-quiet and mutually exclusive from the radio-loud sample.

\begin{figure*} 
    \centering
   \includegraphics[width=.75\textwidth]{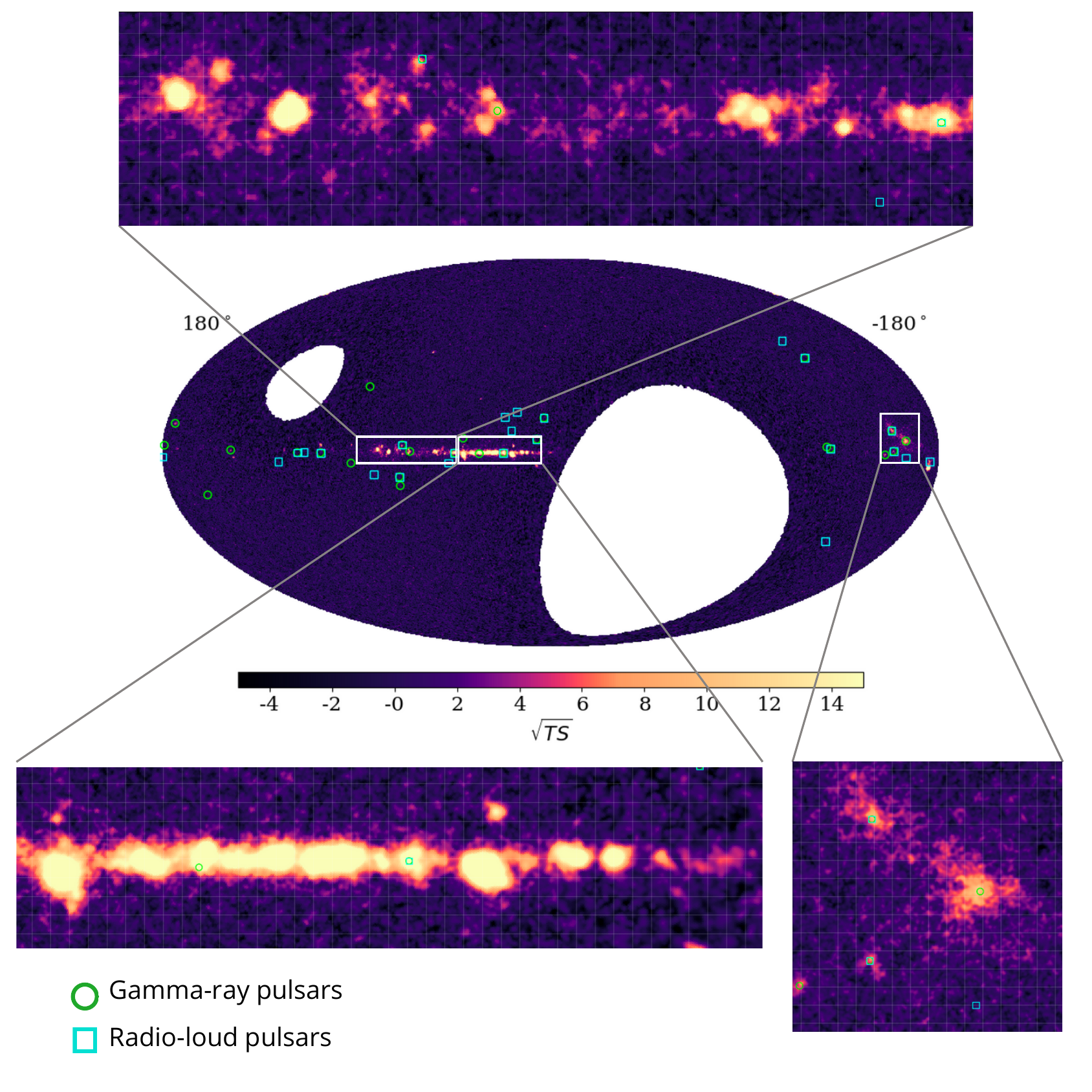}
    \caption{\label{fig:skymap_sigmap} Significance map of the gamma-ray sky observed by HAWC based on 2321 days of data, between 300 GeV and $\sim 100$ TeV. The green circles and cyan squares indicate the positions of the 24 gamma-ray pulsars and 24 radio pulsars used in this analysis, respectively. Among them, 12 pulsars are both gamma-ray-bright and radio-loud.} 
\end{figure*}

Some of the sources are located in highly populated regions of TeV gamma-ray sources. To avoid source confusion, we further removed pulsars within $2^{\circ}.5$ of the TeVCat sources that are not classified as a TeV halo or TeV halo candidate. The choice of $2^\circ.5$ makes sure that the contamination from most TeV sources is avoided while the sample remains sufficiently large.  Our final sample has 36 isolated middle-aged pulsars, including 24 gamma-ray pulsars and 24 radio-loud pulsars (12 pulsars are both radio-loud and gamma-ray bright). Among them, 28 pulsars have not been associated with a known TeV halo or TeV halo candidate. Figure \ref{fig:skymap_sigmap} shows a HAWC significance map of the entire sky above 300 GeV in Galactic coordinates, overlaying the positions of the pulsars used in the analysis. Detailed properties of the pulsars are presented in Appendix.  
 
{\it HAWC Data.---}HAWC is a gamma-ray and cosmic-ray observatory located near Puebla, Mexico at an altitude of 4,100 meters. This analysis uses the HAWC Data Pass~5 \cite{pass5} with 2321 days of exposure. The gamma-ray energy is estimated using the fraction of the photo-multiplier tubes (PMTs) that are triggered in each shower event, called $f_{Hit}$ bins \cite{hawc2017}. 

{\it Stacking Analysis.---}Motivated by the similarity of the observed TeV halos, we combine the observations of the pulsar population accessible to HAWC to constrain the common factors that impact the gamma-ray production in TeV halos. 

We model the TeV halo emission of the i$^{\rm th}$ pulsar in the source list with a spatial model and a spectral model. The spatial model is a Gaussian template with an extension $\sigma$ scaled to the extension of the Geminga TeV halo,  $\sigma_i= {\sigma_{\rm Geminga}} (d_{\rm Geminga} /  d_i)$, where $\sigma_{\rm Geminga}=2^{\circ}$ \cite{Linden17}, $d_{\rm Geminga}=0.25$ kpc and $d_i$ is the distance to the pulsar in kpc. The value of $\sigma_i$ is fixed in the fit. The spectral model describes the flux as a simple power-law, $dN/dEdAdt \propto (E / E_{\rm piv})^{-\alpha}$, where $E_{\rm piv}$ is the pivot energy and $\alpha$ is the spectral index. We fit such a model to data in both the full energy range, 0.316 TeV-100 TeV, and in half-decade bins. In the full energy range the pivot energy is fixed to 30 TeV, while in the fit in each energy bin, the pivot energy is chosen to be the median energy of each energy bin. We fix the spectral index to $\alpha=2.7$ and evaluate the uncertainties associated with the choices of spatial extension and spectral index later in a systematic study. 

We assume that all TeV halos in the population may be described by the same physics, where the differential energy flux of a TeV halo, $F_i \equiv E^2 (dN/dEdAdt)_i$, is scaled to the available power of a pulsar, $A_i$, through a common efficiency, $\eta$: $F_i = \eta\,A_i$. The efficiency $\eta$ represents the fraction of the pulsar's power that is converted into the differential TeV halo luminosity at a given energy. We consider two scenarios for $A_i$, namely, the spin-down flux scenario and the GeV flux scenario. In the former model, $A_i = \dot{E}_i/ (4\pi d_i^2)$, where $\dot{E}_i = 4\pi^2 I \dot{P}_i P_i^{-3}$ is the current-day spin-down power with $I = 10^{45}\,\rm g\,cm^{-2}$ being the moment of inertia of a neutron star which is fixed to be the same for all pulsars. In this scenario, the closer and more energetic pulsars have a higher gamma-ray flux. In the GeV flux scenario, we assume $A_i = F_{0.1-100\,\rm GeV}$, where $F_{0.1-100\,\rm GeV}$ is the integrated flux of the pulsar between 100 MeV and 100 GeV provided by 3PC. In this second scenario, brighter GeV pulsars would produce brighter very-high energy (VHE) gamma-ray halos. Independent of the distance, the spin-down flux and the GeV flux are correlated, though the slope has a large dispersion as shown in \cite{3PC}. We therefore treat them as two separate scenarios in the stacking analysis. 

We fit each source in the pulsar samples using the spectral and spatial model described above. We then add their log-likelihood profiles to constrain the common efficiency $\eta$ in each power scenario and energy bin.  

\begin{figure*} 
    \centering
   \includegraphics[width=.85\textwidth]{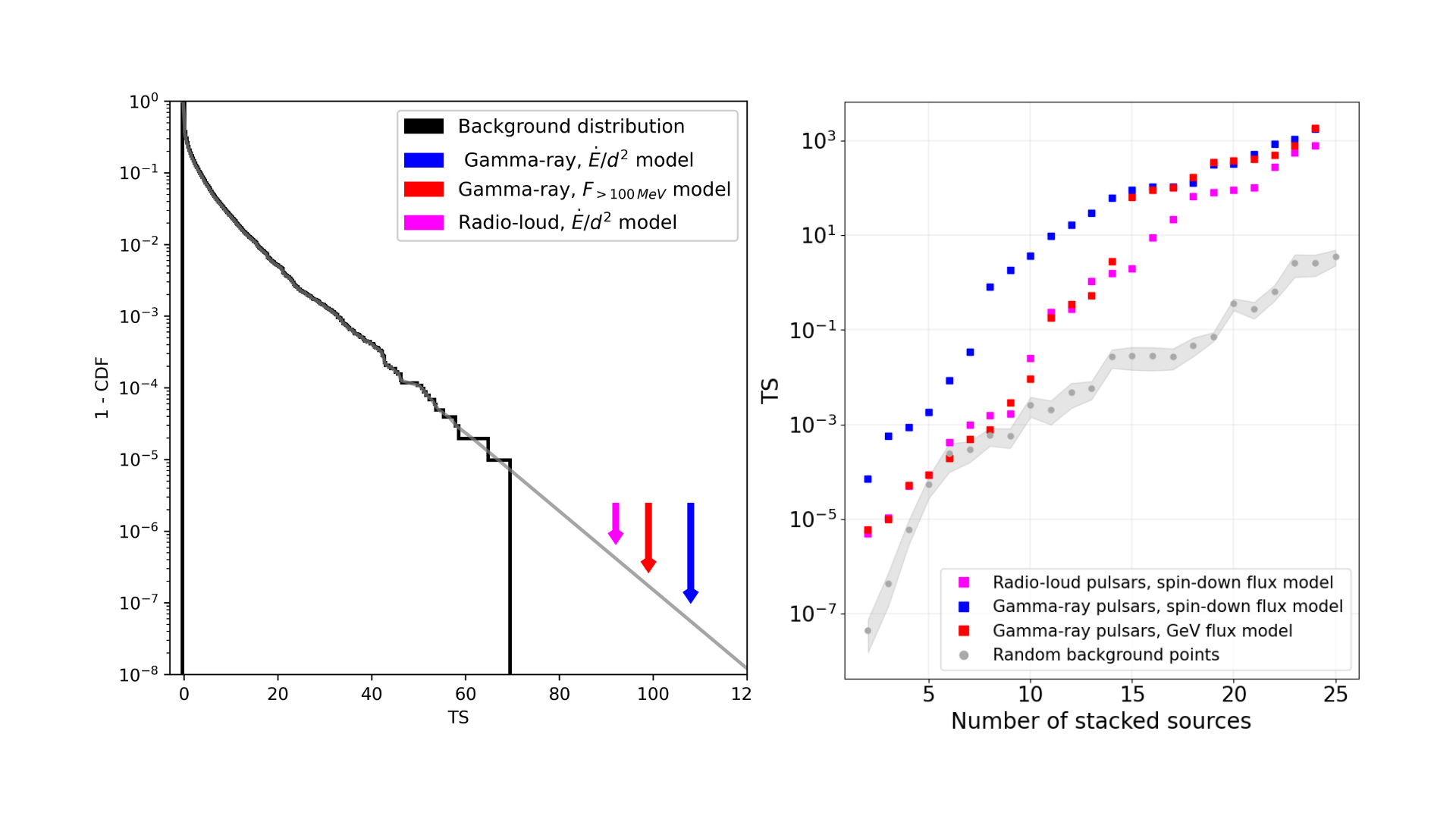}
    \caption{\textbf{Left:} Test statistics (TS) of the stacked likelihoods using radio and gamma-ray pulsars. The black steps indicate the cumulative distribution function of the stacked TS values obtained from samples consisting of random positions. The grey curve is an extrapolation based on the 101,965 trials of background samples. The arrows indicate the results of analyses using different pulsar groups and weighting schemes. \textbf{Right:} TS values resulted from stacking analyses in the full energy range as a function of the number of stacked sources. From right to left, the sources with the highest TS are removed from the sample one by one. Colors indicate the pulsar samples and weighting schemes. The grey shaded region indicates the mean and standard deviation found from the same analysis but with background samples consisting of random positions.}
    \label{fig:TS_dist}
\end{figure*}

We compare the combined signal with a null hypothesis, where the flux of extended TeV halo emission around pulsars is consistent with background fluctuations. 
We first construct a test statistic (TS), which is defined to be twice the logarithm of the ratio of the likelihoods when fitting the data with and without TeV halos around the selected pulsars, ${\rm TS} \equiv 2 \ln [ {\cal L}(\hat{\eta}) / {\cal L} (\eta = 0) ]$. To understand the TS distribution of the null hypothesis, we perform a Monte Carlo simulation to generate random positions based on the spatial distribution of Galactic pulsars (See  Appendix for details regarding the background source generation). These positions are treated as fake sources and we stack their likelihood profiles in the actual data. We use the same selection criteria as for the real sources to form a fake source sample, that is, a fake source needs to be in the HAWC field-of-view and located at least $2^\circ.5$ away from TeVCat sources. The black curve in the left panel of Figure~\ref{fig:TS_dist} presents the distribution of the TS under the null hypothesis. Out of 101,965 trials of stackings of 10 fake sources, the highest TS is found to be ${\mathrm{TS}}_{\rm bg} = 69.5$. Therefore, a ${\mathrm{TS}}$ much higher than  ${\mathrm{TS}}_{\rm bg}$ can safely reject the background hypothesis at least at the level of $4.27\,\sigma$. We verified that the background TS distribution is not sensitively impacted by the number of stacked sources (see Appendix). The background TS distribution departs from a Chi-square over two ($\chi^2/2$) distribution that is expected from the Wilks’ theorem \cite{Wilks:1938dza}. This is caused by the fake sources that land at the outskirts of TeV sources that are more extended than $2^\circ.5$, the separation threshold adopted in our source selection criteria. 

{\it Results.---}We perform the stacking analysis using radio and gamma-ray pulsar samples and the full energy range. We adopt the spin-down flux scenario for radio pulsars, considering that not all radio pulsars are detected in GeV gamma rays. For gamma-ray pulsars, we consider both weighting scenarios. This led to a total of 3 trials of stacking analyses. To avoid bias caused by bright TeV halos, we first consider the sub-sample that is not associated with known TeV halo or TeV halo candidates, which includes 17 gamma-ray and 20 radio-loud pulsars. The highest TS is found to be ${\mathrm{TS}} = 108$, obtained from the trial using gamma-ray pulsars and the spin-down flux scenario. Based on an extrapolation of the simulated background TS distribution (grey curve in left panel of Figure \ref{fig:TS_dist}), this TS value corresponds to a p-value $p = 5.5\times 10^{-8}$, based on an extrapolation of the TS distribution of the background trials.

The other two trials result in ${\mathrm{TS}} = 92$ and $96$, respectively, as also indicated in the figure. We apply a penalty trial factor to account for the fact that multiple trials may increase the possibility of finding a significant event. Although the three signal hypotheses are correlated, we account for a conservative trial factor of 3 when evaluating a post-trials p-value. Specifically, we multiply the lowest p-value by 3 to obtain a post-trial p-value and convert the p-value to a Gaussian significance of $5.10\,\sigma$.
This positive detection of extended TeV gamma-ray emission around pulsar populations not previously associated with TeV halos indicates that TeV halos may commonly exist around middle-aged pulsars.

To ensure that the stacked TS is not dominantly contributed by a few bright sources, we gradually remove the sources with the highest TS values from a pulsar sample and examine the evolution of the stacked TS values. We use all pulsars including those associated with known TeV halos in this test.  The right panel of Figure~\ref{fig:TS_dist} shows that in all weighting scenarios and pulsar groups, the stacked TS gradually declines as the number of stacked sources decreases. The trend cannot be explained by the background which is indicated by the grey shaded region. The evidence of an excess halo emission against the background comes from the entire pulsar population as opposed to just the brightest sources.

Finally, to examine whether the emission is physically extended, we perform the stacking analysis using a point-like spatial template. Similar to the extended-model analysis, we confine the efficiency parameter using a combined likelihood. We compare the two spatial models based on the TS values and the Bayesian Information Criterion (BIC) from the analysis using the full energy range. We find that in all weighting scenarios and pulsar groups, the extended model is significantly preferred \cite{bic} over the point-like model with ${\rm TS}_{\rm extended}-{\rm TS}_{\rm point} >25$ and ${\rm BIC}_{\rm point} - {\rm BIC}_{\rm extended}>10$.

The halo efficiencies computed using all pulsars are generally consistent with those from pulsars not associated with known TeV halos within uncertainties as shown in Figure~\ref{fig:efficiencies} in Appendix. The systematic uncertainties include the effects due to spectral models, extension models, as well as uncertainties in our detector configurations. See Appendix for details regarding the evaluation of the systematic errors. The efficiencies decrease as a function of the photon energy because the gamma-ray spectrum follows a power-law distribution. On average, the differential energy flux in TeV halos is at the level of $(0.01-1)\%$ of the spin-down flux. For gamma-ray pulsars, the energy deposited in TeV halos ranges from $(0.1-10)\%$ of that in the GeV emission of the pulsars.

{\it Implications of Commonly existing TeV halos.---}We have used a stacking technique that combines the likelihood profiles of  radio and gamma-ray pulsars accessible to HAWC to test the hypothesis that the flux of extended TeV halo emission around pulsars is consistent with background fluctuations. Our analysis has assumed that the TeV flux of a pulsar halo scales with the spin-down flux or the GeV gamma-ray flux of the pulsar. Using pulsars that have not been associated with a known TeV halo, the null hypothesis is   rejected at the level of $5.10 \,\sigma$. 

Our analysis implies an existence of ubiquitous TeV halos around isolated middle-aged pulsars. This establishes TeV halos as a new pulsar phenomenon that has a distinct origin from pulsar wind nebulae, as the latter are usually orders of magnitude more compact for pulsars at this age. 

A common existence of TeV halos suggests that the production mechanism of TeV halos is more likely related to the transport of relativistic particles in the vicinity of a pulsar than the local environment of the observed halos.
It also suggests a leptonic origin of the TeV gamma-ray emission since hadronic emission highly depends on the gas density (e.g.,\cite{2023PhRvD.107l3020S}) and would vary from pulsar to pulsar. 

The longer confinement of relativistic electrons by the halos than the average ISM has strong implications for indirect searches for dark matter. While these halos provide an alternative way to explain the GeV excess in the Galactic center \cite{2018PhRvD..98d3005H}, they also make it harder to explain the cosmic-ray positron excess \cite{PAMELA:2008gwm, 2012PhRvL.108a1103A, PhysRevLett.122.041102} with nearby pulsars \cite{gemingahawc17,2020PhRvD.102b3015M, 2023PhRvD.107l3020S}, compared to scenarios involving exotic physics. A common existence of TeV halos also impacts the view of Galactic gamma-ray astronomy. These halos may contribute to the diffuse Galactic plane emission observed by Tibet AS$\gamma$ \cite{Tibet21}, HAWC \cite{HAWC:2021bvb}, and LHAASO  \cite{lhaasoGDE} experiments and explain the hardening of the diffuse $\gamma$-ray emission around 1~TeV \cite{2018PhRvL.120l1101L,2022ApJ...928...19V, 2023arXiv230600051D, 2023arXiv230617275F}. 

Our study suggests that a conversion of the spin-down power to the TeV halo emission could be as efficient as $\sim 0.1\%$ and that the power carried by a TeV halo is about $1\%$ of that by the GeV magnetospheric emission at 10~TeV (See Figure \ref{fig:efficiencies}). 
The spin-down flux model resulted in a higher TS than the GeV flux model, though the difference is not statistically significant. This could be related to the fact that the  observation of GeV magnetospheric emission may be biased by orientation effects. We further extended this analysis to constrain the diffusion radius of electrons around our sample of TeV halos. From this, we derived a diffusion coefficient of $D_0\approx 2.0\times 10^{27}\,\rm cm^2\,s^{-1}$ at a reference energy of $E_0 = 10$~GeV, consistent with measurements around Geminga \cite{gemingahawc17} (see Appendix and Table \ref{tab:diffusion-results} for more details regarding the diffusion coefficient study). These results reinforce the idea that electron propagation in TeV halos occurs in a suppressed diffusion environment, significantly lower than the typical Galactic diffusion coefficient. 

Future observations of the spatial profiles of individual halos over a wide energy range would be needed to understand how electrons diffuse in TeV halos. Wide-field gamma-ray detectors with better sensitivity and covering the Southern sky can carry out such population studies \cite{2019BAAS...51g.109H}. Finally, while pulsars are typically identified via their pulsed emission in radio, X-ray, and GeV gamma-ray wavelengths, TeV halos open a new window on pulsar observations. In particular, they offer a unique way to probe the ``invisible" pulsars \cite{Linden17} with beaming angles misaligned with observers and middle-aged and old pulsars with a surface density too low to be detected in other wavelengths. 

\vspace{2em}

\begin{acknowledgments}
We acknowledge the support from: the US National Science Foundation (NSF); the US Department of Energy Office of High-Energy Physics; the Laboratory Directed Research and Development (LDRD) program of Los Alamos National Laboratory; Consejo Nacional de Ciencia y Tecnolog\'{i}a (CONACyT), M\'{e}xico, grants LNC-2023-117, 271051, 232656, 260378, 179588, 254964, 258865, 243290, 132197, A1-S-46288, A1-S-22784, CF-2023-I-645, c\'{a}tedras 873, 1563, 341, 323, Red HAWC, M\'{e}xico; DGAPA-UNAM grants IG101323, IN111716-3, IN111419, IA102019, IN106521, IN114924, IN110521 , IN102223; VIEP-BUAP; PIFI 2012, 2013, PROFOCIE 2014, 2015; the University of Wisconsin Alumni Research Foundation; the Institute of Geophysics, Planetary Physics, and Signatures at Los Alamos National Laboratory; Polish Science Centre grant, 2024/53/B/ST9/02671; Coordinaci\'{o}n de la Investigaci\'{o}n Cient\'{i}fica de la Universidad Michoacana; Royal Society - Newton Advanced Fellowship 180385; Gobierno de España and European Union -
NextGenerationEU, grant CNS2023- 144099; The Program Management Unit for Human Resources \& Institutional Development, Research and Innovation, NXPO (grant number B16F630069); Coordinaci\'{o}n General Acad\'{e}mica e Innovaci\'{o}n (CGAI-UdeG), PRODEP-SEP UDG-CA-499; Institute of Cosmic Ray Research (ICRR), University of Tokyo. H.F. acknowledges support by NASA under award number 80GSFC21M0002. C.R. acknowledges support from National Research Foundation of Korea (RS-2023-00280210). We also acknowledge the significant contributions over many years of Stefan Westerhoff, Gaurang Yodh and Arnulfo Zepeda Dom\'inguez, all deceased members of the HAWC collaboration. Thanks to Scott Delay, Luciano D\'{i}az and Eduardo Murrieta for technical support.

\end{acknowledgments}

\newcommand {\aap}      {A\&A}
\newcommand {\aj}       {Astronomical Journal}
\newcommand {\apjl}     {ApJ Letters}
\newcommand {\apjs}     {The Astrophysical Journal Supplement Series}
\newcommand {\araa}     {Annual Review of Astronomy and Astrophysics}
\newcommand {\aapr}     {A\&A Reviews}
\newcommand {\jcap}     {JCAP}
\newcommand {\physrep}  {Physics Reports}
\newcommand {\mnras}    {MNRAS}
\newcommand {\ssr}      {Space Science Reviews}

\bibliography{references.bib}

\clearpage 
\appendix
\section{Appendices}
\section{Properties of Pulsars}
\begin{figure}
    \centering
    \includegraphics[width=0.49\textwidth]{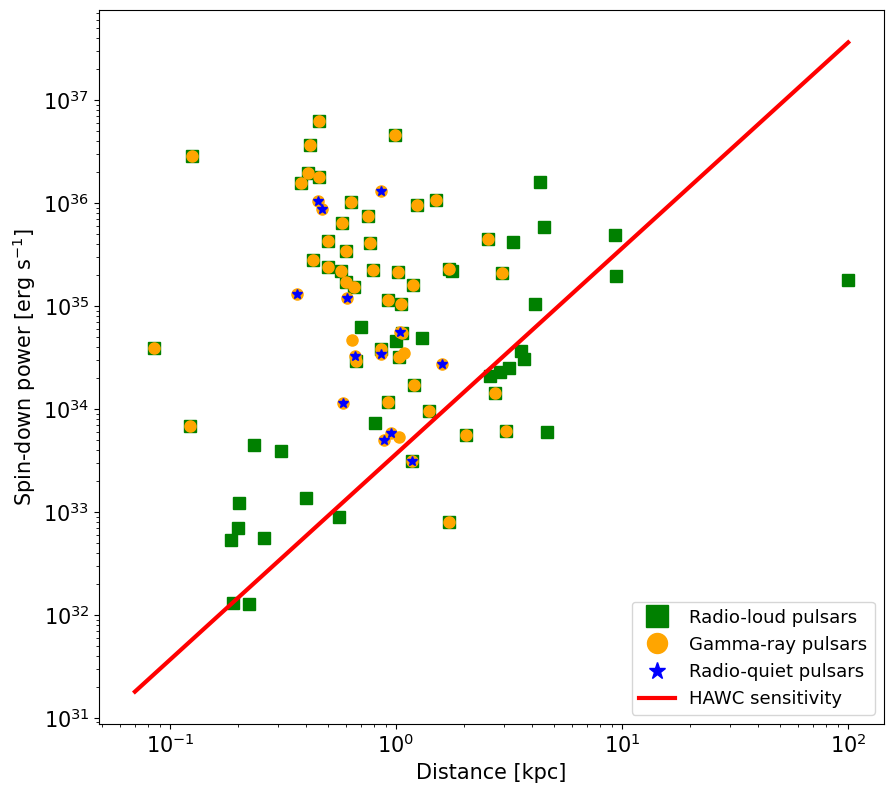}
    \caption{Detection space of isolated pulsars. The green squares represent radio-loud pulsars, the orange dots correspond to gamma-ray pulsars (detected by \textit{Fermi}-LAT), the blue stars correspond to radio-quiet pulsars, and the red line corresponds to the pass 5 HAWC sensitivity at 7~TeV.}
    \label{fig:dist-Edot}
\end{figure}

Figure \ref{fig:dist-Edot} shows the current-day spin-down power and distances of the isolated middle-aged pulsars that are in the HAWC's field-of-view. The red line indicates the sensitivity, which is estimated as follows. At 7 TeV, for an $E^{-2.5}$ spectrum at declination $\delta= 22^{\circ}$ where the detector is the most sensitive, the HAWC sensitivity \cite{3HWC} is 
\begin{equation} \label{eq:hawc-sensi}
    \frac{dN}{dE} = 5\times 10^{-14}\, \mathrm{TeV}^{-1}\mathrm{cm}^{-2}\mathrm{s}^{-1}\left(\frac{E}{7\,\mathrm{TeV}}\right)^{-2.5}. 
\end{equation}
Based on this value, we then applied the $\dot{E}/4\pi d^2= 2.91\times 10^{32}\,\mathrm{erg}\,\mathrm{kpc}^{-2}\mathrm{s}^{-1}$ cut to select the pulsars. Along with the optimal HAWC sensitivity, this cut uses a generous 5\% TeV halo efficiency  out of the spin-down power to include as many potential emitters as possible.

\section{Background Generation}
The spatial distribution of pulsars in the Galaxy may be described by \cite{Steppa:2020qwe}
\begin{equation}\label{eqn:rho_pulsar}
    \rho(r, z) = \left(\frac{r + r_{\rm off}}{R_\odot + r_{\rm off}}\right)^\alpha \exp\left[-\beta \left(\frac{r - R_\odot}{R_\odot + r_{\rm off}}\right)\right]  \exp\left(-\frac{|z|}{z_0}\right)
\end{equation}
where $r$ and $z$ are the distance to the Galactic center and height over the Galactic plane, respectively. We take $R_\odot = 8.5$~kpc, $r_{\rm off} = 0.55$~kpc, $\alpha = 1.64$, $\beta = 4.01$, and $z_0 = 0.18$~kpc \cite{Steppa:2020qwe}. We have assumed that the distribution is azimuth-symmetric, that is, $\rho$ does not depend on the azimuthal angle $\phi$.

\begin{figure*}
    \centering
    \includegraphics[width=0.85\textwidth]{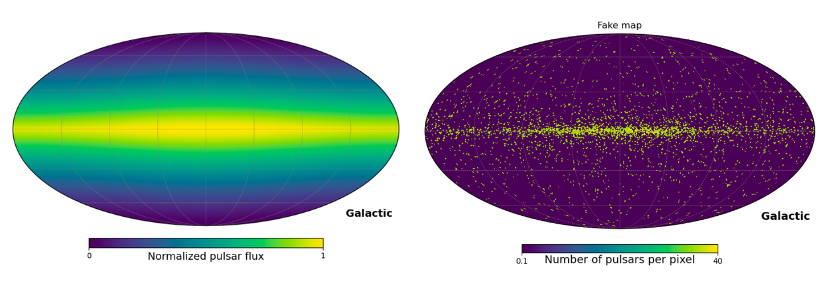} 
    \caption{\textbf{Left:} Pulsar flux per solid angle per area based on the spatial  distribution of Galactic pulsars. Calculated from equation \ref{eqn:N_bl}. \textbf{Right:} Random sources that are generated using a density map after normalizing the pulsar flux map in the left panel. Both maps are in Galactic coordinates.}
    \label{fig:pulsar_distrubution_model}
\end{figure*}

The expected emission from pulsars in the direction given by of Galactic  longitude $l$ and latitude $b$ per solid angle $d\Omega = \sin(b) db dl$ per area, assuming that all pulsars are equally luminous, is obtained by integrating the three-dimensional pulsar density over the line of sight: 
\begin{equation}\label{eqn:N_bl}
    \frac{dN}{d\Omega dA} \propto \int_0^\infty ds\, \rho (s, l, b)    
\end{equation}
where $s$ is the line of sight (i.e., distance to the observer). The pulsar density at position $(s, l, b)$ in the Galactic coordinate is obtained through equation~\ref{eqn:rho_pulsar} at the corresponding position in the Galactocentric coordinate $(r, \phi, z)$ (see the left panel of Figure \ref{fig:pulsar_distrubution_model}). 

Using equation~\ref{eqn:N_bl}, we create a healpy map for every direction $(l, b)$. We normalize the map by its maximum pixel, yielding a probability map of seeing a pulsar in a random direction. We generate random source positions using this probability map. We verified that the spatial maps of fake sources generated through this process (see the right panel of Figure \ref{fig:pulsar_distrubution_model}) and actual pulsars in the ATNF catalog  are consistent.


\section{Background Test Statistic Distribution}
\begin{figure}
    \centering
    \includegraphics[width=0.49\textwidth]{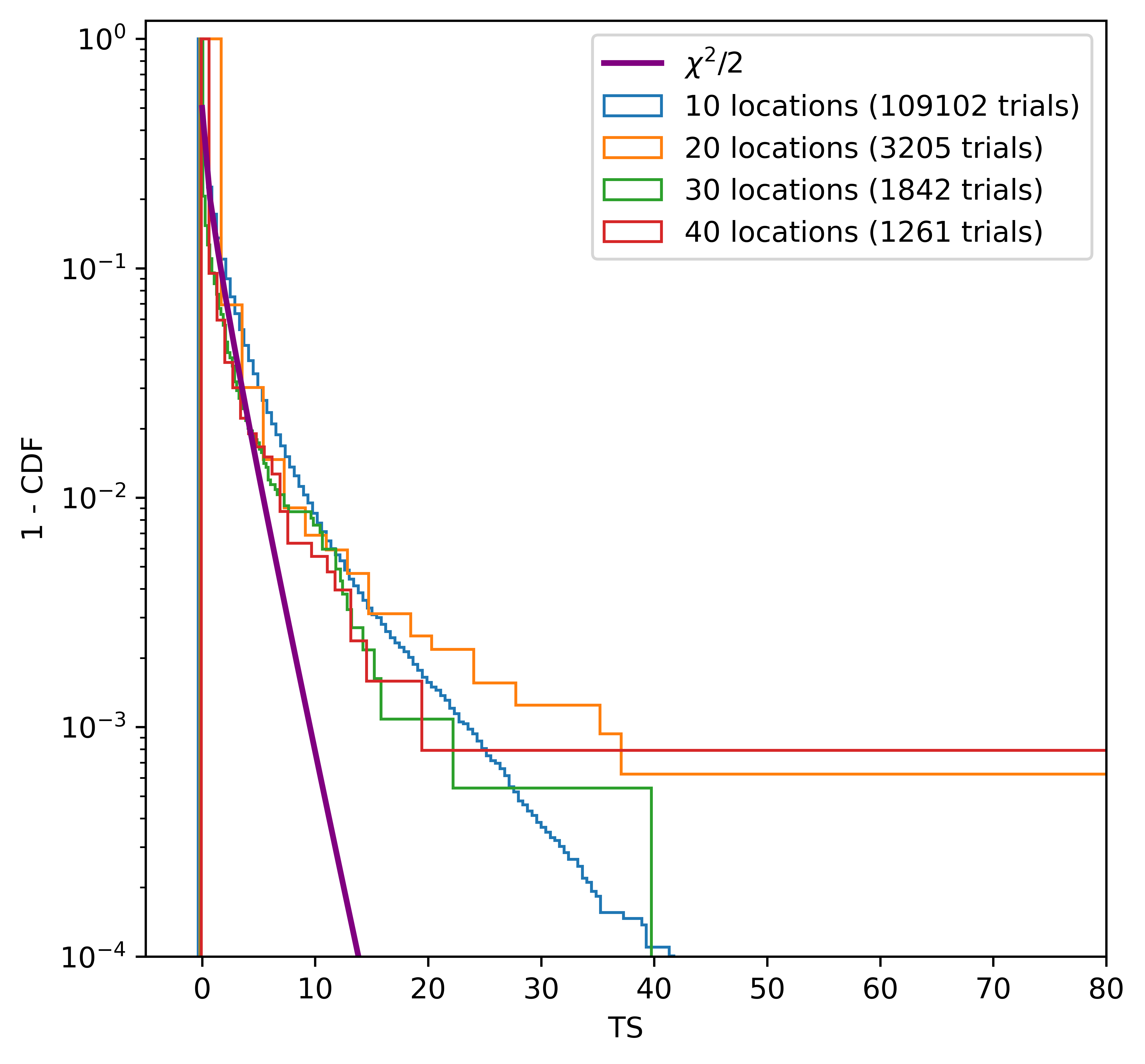}
    \caption{TS distribution for 10, 20, 30 and 40-source sample of random locations. The $\chi ^2/2$ distribution shape remains the same since the only free parameter is the normalization flux.}
    \label{fig:TS_dist4in1}
\end{figure}

We perform stacking analyses using the background trials generated following the algorithm described in the last section. Due to computational limits, we stack 10 fake sources, instead of $\sim 20$ sources in the real sample, for $\sim 10^5$ times. We first verify that the background TS distribution does not have a strong dependence on the number of stacked sources. 
Figure~\ref{fig:TS_dist4in1} shows the TS distributions when stacking 10, 20, 30 and 40 fake sources. The curves for 20 and above random locations are computed with less trials due to long computation time. They present similar shapes and agree with the distribution found with 10 locations. All cases present a tail that departs from the $\chi^2/2$ distribution. In the context that the TS values follows a $\chi^2$ distribution asymptotically, the term $\chi^2/2$ arises as only a  positive normalization of the gamma-ray flux is allowed in the fits. 

Beyond the distribution of the $\sim 10^5$ simulated background trials, we evaluated the TS by performing a linear extrapolation of the tail of simulated distribution. Other extrapolation methods such as polynomial fits fail to provide a reliable extrapolation to the range of the TS values obtained by our analysis.

We note the caveat that HAWC's field-of-view is smaller than the total sky area covered by the fake sources of the $\sim10^5$ trials. This implies that the trials are not entirely independent.

\section{Likelihood analysis}
All the analysis is performed using the HAWC Accelerated Likelihood (HAL) plugin to the Multi-Mission Maximum Likelihood Framework (3ML) \cite{3ml,hal}. Gamma-hadron selection is applied to the data, which is divided into nine bins based on the $f_{hit}$ parameter space. The $f_{hit}$ parameter, defined as the ratio of triggered PMTs to the total number of active PMTs in the array, serves as a proxy for energy, with lower $f_{hit}$ values corresponding to lower-energy gamma rays. In this work, the energy range covered goes from 0.316-100 TeV, divided into 5 half-decade energy bins (see Table \ref{tab:energy-bins}).
\begin{table}
    \centering
    \caption{Reconstructed energy bins used for the analysis.}
    \begin{tabular}{ccc}
    \hline
        Bin & Energy range [TeV] & Median energy [TeV]\\
        \hline
        a & 0.316-1.00 & 0.56\\
        b & 1.00 - 3.16 & 1.78\\
        c & 3.16 - 10.0 & 5.62\\
        d & 10.0 - 31.6 & 17.8\\
        e & 31.6 -100 TeV & 56.2\\
    \hline
    \end{tabular}
    \label{tab:energy-bins}
\end{table}
\section{Halo Efficiencies}
\begin{figure*}[t]
    \centering
    \includegraphics[width=.49\textwidth]{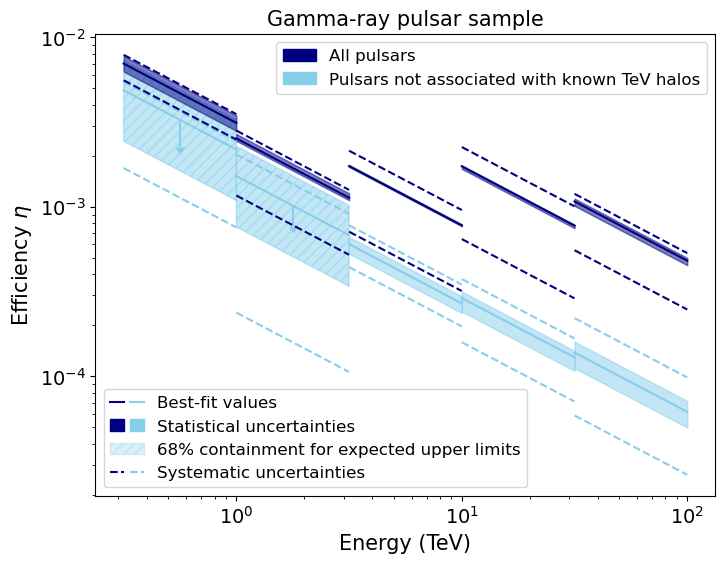}
    \includegraphics[width=.49\textwidth]{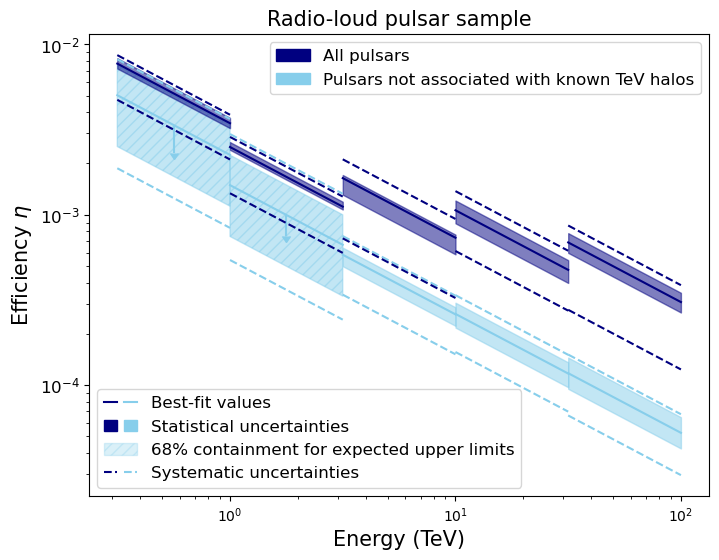}
    \includegraphics[width=.49\textwidth]{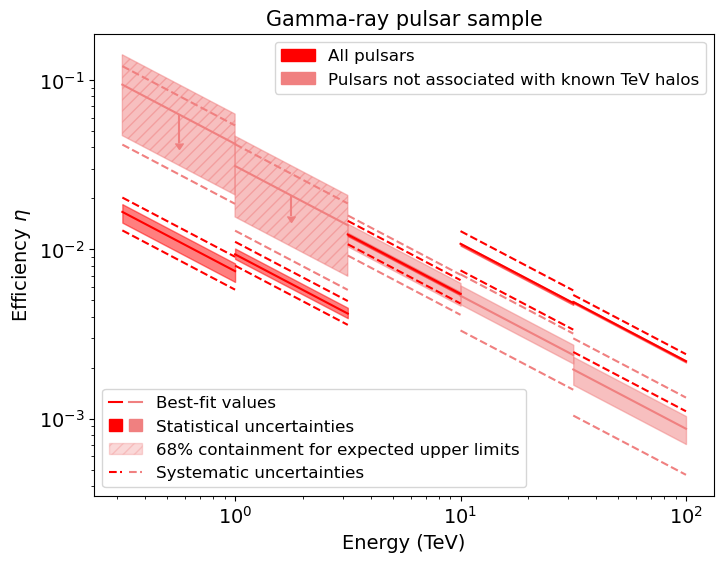}
    \caption{\label{fig:efficiencies} The halo efficiency as a function of energy. \textbf{Upper}: spin-down power scenario for gamma-ray (left) and radio-loud (right) samples, respectively. \textbf{Bottom}: GeV flux scenario for the gamma-ray pulsar sample. The results computed using the full source sample including pulsars associated with known TeV halo and TeV halo candidates are shown in dark colors. Those from the sub-sample not previously associated with TeV halos or TeV halo candidates are shown in light colors. The measurements are indicated as solid curves. The statistical and systematic uncertainties are indicated as shaded bands and dashed lines, respectively. In the first two energy bins where ${\rm TS}<25$, the 68\% containment regions of the expected upper limits at the 90\% confidence levels calculated using a Poisson-fluctuated background are shown as hatched bands.}
\end{figure*}

Figure \ref{fig:efficiencies} presents the common efficiency of the TeV halos, $\eta$, in each energy bin obtained using the spin-down flux scenario (blue; with both gamma-ray and radio-loud pulsars) and GeV flux scenario (red; with gamma-ray pulsars).  
\section{Systematic Uncertainties} Tables \ref{tab:sys-modelB} and \ref{tab:sys-modelC}  summarize the uncertainties of the halo efficiency measurements due to the following factors in five energy bins. 
\begin{table*}
\caption{Systematic uncertainties for both pulsar samples in the spin-down power scenario. The percentages correspond to  variations of the efficiency from the nominal values.}
\label{tab:sys-modelB}
\begin{ruledtabular}
\begin{tabular}{ccc}
Energy range [TeV] & With known TeV halos & Without known TeV halos \\
\hline
\multicolumn{3}{c}{Radio-loud pulsars} \\
\hline
0.316-1.00 & -39\% , +12\% & -45\% , +18\%\\
1.00-3.16 & -46\% , +15\% & -52\% , +64\%\\
3.16 - 10.0 & -56\% , +29\% & -42\% , +29\%\\
10.0 - 31.6 & -42\% , +29\% & -40\% , +29\%\\
31.6 - 100 & -60\% , +26\% & -43\% , +28\%\\
\hline
\multicolumn{3}{c}{Gamma-ray pulsars} \\
\hline
0.316-1.00 & -20\% , +13 \% & -50\% , +62\%\\
1.00-3.16 & -54\% , +11\% & -79\% , +77\%\\
3.16 - 10.0 & -59\% , +23\% & -27\% , +29\%\\
10.0 - 31.6 & -63\% , +30\% & -45\% , +29\%\\
31.6 - 100 & -48\% , +11\% & -57\% , +60\%\\
\end{tabular}
\end{ruledtabular}
\end{table*}

\begin{table*}
\caption{Systematic uncertainties for the gamma-ray pulsar sample in the GeV power scenario. The percentages correspond to the variations of the efficiency from the nominal values.}
\label{tab:sys-modelC}
\begin{ruledtabular}
\begin{tabular}{ccc}
Energy range [TeV] & With known TeV halos & Without known TeV halos \\
\hline
0.316-1.00 & -22\% , +21\% & -35\% , +78\%\\
1.00-3.16 & -14\% , +18\% & -45\% , +77\%\\
3.16 - 10.0 & -12\% , +20\% & -26\% , +28\%\\
10.0 - 31.6 & -30\% , +19\% & -38\% , +34\%\\
31.6 - 100 & -49\% , +9\% & -46\% , +53\%\\
\end{tabular}
\end{ruledtabular}
\end{table*}
{\it Detector systematics.---}We investigate how the results change when using different detector configurations, such as variations in charge measurement and efficiency of the PMTs, as well as the arrival time distribution at the PMTs. For more details about these sources of systematic uncertainties, see \cite{hawc2017}. The relative contribution of the detector systematic varies with energy from 35\% to 68\% for the radio loud sample and from 18\% to 82\% for the gamma-ray sample.

{\it Spectral Model systematics.---}We run the entire analysis assuming three different scenarios: spectral index of -2.0, -3.0, and a power law with an exponential energy cutoff with $E_{\rm cut}=60$ TeV and spectral index of -2.7. The relative contribution of the chose of spectral index varies with energy from 6\% to 54\% for the radio loud sample and from 8\% to 48\% for the gamma-ray sample. And the relative contribution of the chose of spectral shape varies with energy from 5\% to 49\% for the radio loud sample and from 5\% to 43\% for the gamma-ray sample.

{\it Extension systematics.---}We perform the entire analysis using a single value for the extension, namely, 0.48 degrees. This value is the average of the expected extensions from scaling Geminga's physical halo size by the pulsar distance. For  sources with known TeV halos, we use the reported extension in the literature. The relative contribution  due to the choice of one single extension value varies with energy from 22\% to 42\% for the radio loud sample and from 17\% to 58\% for the gamma-ray sample.

{\it Age threshold.---}We remove pulsars younger than 100 kyr from the source list. For the gamma-ray sample, we remove 5 pulsars (from which one pulsar is a TeV halo candidate), and for the radio-loud sample, we remove 3 pulsars (none of them is a known TeV halo or a TeV halo candidate). The TS values for the sub-samples without TeV halos are still above the background $\mathrm{TS}$, ${\mathrm{TS}}_{\rm bg} = 69.5$. For the gamma-ray sample, we find a $\mathrm{TS}=80.74$ and $\mathrm{TS}=75.9$ with the spin-down power and GeV scenarios, respectively. For the radio-loud sample, we obtain $\mathrm{TS}=80$ with the spin-down power scenario. The relative contribution due to setting the pulsar age threshold to 100 kyr varies with energy from 23\% to 62\% for the radio loud sample and from 13\% to 52\% for the gamma-ray sample.

\section{Halo Size and Diffusion Coefficient}
To use our data to confine the diffusion coefficient of the TeV halos, we further performed a two-dimensional stacking on the halo's gamma-ray efficiency $\eta$ and electron diffusion radius $r_d$. We assume that the gamma-ray flux of a source $i$ at angular distance $\theta$ from the pulsar follows $F_i (\theta) = \eta A_i f_i  (\theta)$. The function $f_i (\theta)$ describes the morphology of the gamma-ray flux such that $\int f_i(\theta) 2\pi \sin \theta d\theta = 1$. Th function may be obtained by integrating the electron distribution over the line-of-sight \cite{gemingahawc17}, 
\begin{equation}
    f_i  (\theta) = \frac{1.22}{\pi^{3/2}\theta_d (\theta + 0.06\,\theta_d)}\exp\left(-\frac{\theta^2}{\theta_d^2}\right),
\end{equation}
where $\theta_d\equiv r_d / d_i$ is the angular size corresponding to the diffusion radius of electrons, $r_d \equiv 2 \left (D(E_e) t(E_e)\right)^{1/2}$ with $D(E_e)$ and $t(E_e)$ being the diffusion coefficient and cooling time of electrons of energy $E_e$. As in \cite{gemingahawc17} we take $E_e = 100$~TeV as it corresponds to the typical gamma-ray energy detected by HAWC. The cooling time is computed assuming that electrons loss energy by sychrotron radiation in the average interstellar medium magnetic field of $3\,\mu$G and inverse Compton scattering the cosmic microwave background. The obtained $D(E_e)$ is then converted to a diffusion coefficient $D_0$ at a reference energy of $E_0=10$~GeV through  $D(E_e) = D_0 (E_e / E_0)^\delta$ with $\delta = 1/3$.

Table \ref{tab:diffusion-results} presents the efficiency, diffusion radius, and diffusion coefficient for the different pulsar samples analyzed in this work. The best-fit $r_d$ corresponds to a diffusion coefficient of $\sim2\times 10^{27}\,{\rm cm}^2 {\rm s}^{-1}$ for 100 TeV electrons,  consistent with the findings of individual halos \citep{gemingahawc17}. The statistical uncertainty of the diffusion radius can be large due to the contamination of nearby sources. These results point to a reduced diffusion in the halos and a more concentrated gamma-ray emission, as electrons are less likely to escape the region where they were injected. 

\begin{table*}
\caption{Efficiency $\eta$, diffusion radius $r_d$ and diffusion coefficcient $D_0$  in the 0.32-100 TeV energy range.}
\label{tab:diffusion-results}
\renewcommand{\arraystretch}{1.5}
\begin{ruledtabular}
\begin{tabular}{lcccc}
Sample & TS & $\eta$ & $r_d$ & $D_0$\\
 & & & [pc] & [${\rm cm}^2/s$]\\
\hline
\multicolumn{5}{c}{All pulsars} \\
\hline
Radio-loud & 1954 & $(1.67^{+0.17}_{-0.42})\times 10^{-3}$ & $54.1^{+234}_{-18.8}$ &  $(2.21^{+60.5}_{-1.27})\times 10^{27}$ \\
Gamma-ray & 3675 & $(1.35\pm 0.06)\times 10 ^{-3}$  & $52.5^{+35.4}_{-4.29}$ & $(2.08^{+3.75}_{-0.33})\times 10^{27}$\\
\hline
\multicolumn{5}{c}{Without known TeV halos} \\
\hline
Radio-loud & 341 & 
$(9.16^{+0.96}_{-1.53})\times 10^{-4}$ & $52.5^{+28.6}_{-18.1}$ & $(2.08^{+2.89}_{-1.19})\times 10^{27}$\\

Gamma-ray & 347 & $(1.41^{+0.17}_{-0.22})\times 10^{-3}$  & $52.5^{+360}_{-7.31}$ & $(2.08^{+126}_{-0.54})\times 10^{27}$\\
\end{tabular}
\end{ruledtabular}
\end{table*}

\section{List of Sources}
Table \ref{tab:sample} lists the pulsars along with their physical information used in the analysis.
\begin{table*}
\caption{Pulsar sample.\label{tab:sample}}
\label{tab:my_label}
\begin{ruledtabular}
    \begin{tabular}{| l | c | c | c | c | c | c | c | c | c | c |}
         PSR & RA & Dec & $l$ & $b$ & $\dot{E}$ & Age & Distance & $\rm{Flux}_{>100\rm MeV}$ \\
         & [$\circ$] & [$\circ$] & [$\circ$] & [$\circ$] & [$\times 10^{35}\,\rm erg\cdot s^{-1}$] & [kyr] & [kpc] & [$\times 10^{-11}\rm erg\cdot cm^{-2}\cdot s^{-1}$]\\
         \hline
        J0002+6216 & 0.742 & 62.27 & 117.3  & -0.074 & 1.53 & 306.3 & 6.357 & 1.88 \\
        B0114+58   & 19.41 & 59.24 & 126.3  & -3.457 & 2.21 & 274.9 & 1.768 & - \\
        J0357+3205 & 59.47 & 32.09 & 162.8  & -16.01 & 0.06 & 537.3 & 0.835 & 6.01 \\
        J0359+5414 & 59.86 & 54.25 & 148.2  & 0.883  & 13.17& 75.27 & 3.45  & 1.98 \\
        B0450-18   & 73.14 & -17.99& 217.1  & -34.087 & 0.01 & 1511.0& 0.4   & - \\
        J0538+2817 & 84.60 & 28.29 & 179.7 & -1.686 & 0.49 & 617.1 & 1.3   & - \\
        B0540+23   & 85.79 & 23.48 & 184.4 & -3.318 & 0.41 & 252.7 & 1.565 & - \\
        J0554+3107 & 88.52 & 31.13 & 179.1  & 2.697 & 0.56 & 51.66 & 1.9   & 1.87 \\
        J0611+1436 & 92.83 & 14.61 & 195.4 & -2.004& 0.07 & 1160 & 0.807  & - \\
        J0622+3749 & 95.54 & 37.82 & 175.9 & 10.96 & 0.27 & 207.7 & 1.6    & 1.77 \\
        J0631+1036 & 97.86 & 10.62 & 201.2 & 0.450 &1.70 & 44.39 & 2.105 & 3.03 \\
        J0633+0632 & 98.43 & 6.542 & 205.1 & -0.933& 1.19 & 59.21 & 1.355 & 9.56 \\
        J0633+1746 & 98.48 & 17.77 & 195.1 & 4.266 & 0.33 & 342.3 & 0.25 & 422.10 \\
        B0656+14   & 105.0 & 14.24 & 201.1 & 8.258 & 0.38 & 111.0 & 0.288 & 2.65 \\
        J0729-1448 & 112.3 & -14.81& 230.4 & 1.421 & 2.81 & 35.18 & 2.681 & 0.60 \\
        J0734-1559 & 113.7 & -15.99& 232.1 & 2.015 & 1.32 & 196.5 & 1.3 & 4.58 \\
        B0919+06   & 140.6 & 6.639 & 225.4 & 36.39 & 0.07 & 498.1 & 1.1 & 0.22 \\
        B0950+08   & 148.3 & 7.927 & 228.9 & 43.69 & 0.01 & 17440.0 & 0.261 & - \\
        B1702-19   & 256.4 & -19.11& 3.188 & 13.03 & 0.06 & 1146.0 & 0.747 & 0.36 \\
        J1725-0732 & 261.3 & -7.55 & 15.87 & 15.32 & 0.01 & 8848.0 & 0.203 & - \\
        J1741-2054 & 265.5 & -20.9 & 6.422 & 4.906 & 0.09 & 385.6 & 0.3 & 11.86 \\
        B1740-03   & 265.8 & -3.653& 21.65 & 13.40 & 0.01 & 4487.0 & 0.2 & - \\
        J1755-0903 & 268.8 & -9.064& 18.32 & 8.150 & 0.04 & 3881.0 & 0.235 & - \\
        J1831-0952 & 277.9 & -9.867& 21.89 & -0.128& 10.8 & 128.2 & 3.863 & 5.44 \\
        J1836+5925 & 279.1 & 59.43 & 88.87 & 24.99 & 0.11 & 1827.0 & 0.3 & 61.95 \\
        J1846+0919 & 281.6 & 9.329 & 40.69 & 5.342 & 0.34  & 359.8 & 1.53 & 3.57 \\
        J1913+1011 & 288.3 & 10.19 & 44.48 & -0.167& 28.9  & 168.2 & 4.62 & 1.83\\
        B1929+10   & 293.1 & 10.99 & 47.38 & -3.885& 0.04 & 3106.0 & 0.31 & - \\
        B1951+32   & 298.2 & 32.88 & 68.76 & 2.823 & 37.23 & 107.5 & 3.0 & 14.77 \\
        J1954+2836 & 298.6 & 28.60 & 65.24 & 0.377 & 10.5 & 694.23 & 1.96 & 10.7 \\
        J1957+5033 & 299.4 & 50.56 & 84.58 & 11.01 & 0.05 & 838.2 & 1.365 & 2.62 \\
        J2043+2740 & 310.9 & 27.68 & 70.61 & -9.151& 0.55 & 1231.0 & 1.48 & 0.90 \\
        J2055+2539 & 314.0 & 25.67 & 70.69 & -12.52& 0.05 & 1234.0 & 0.62 & 5.32 \\
        J2116+3701 & 319.1 & 37.02 & 82.30 & -8.341& 0.67 & 439 & 2.903 & - \\
        J2139+4716 & 325.0 & 47.27 & 92.63 & -4.019& 0.03 & 2517.0 & 0.8 & 2.70 \\
        J2240+5832 & 340.2 & 58.54 & 106.5 & -0.111& 2.20 & 145.4 & 7.275 & 0.98 \\
        B2334+61   & 354.3 & 61.85 & 114.3 & 0.233 & 0.63 & 40.54 & 0.7 & - \\
    \end{tabular}
\end{ruledtabular}
\end{table*}
\end{document}